# GUIMesh: a tool to import STEP geometries into Geant4 via GDML

M. Pinto[a]*, P. Gonçalves[a]

[a]*LIP-Lisboa, Av. Gama Pinto, n.2, piso 3, 1649-003 Lisboa, Portugal*

**Abstract**

Detailed radiation analysis of instruments flown in space is critical to ensure mission safety, often requiring the use of state-of-the-art particle transport simulation tools. Geant4 is one of the most powerful toolkits to simulate the interaction and the passage of particles through matter. This framework however, is not prepared to receive Standard for The Exchange of Product data (STEP) files, the most versatile Computer-Aided Design (CAD) format, as inputs requiring previous conversion to other formats. This, especially when the instruments have complex shapes, and/or a large number of volumes may lead to loss of detail and under or overestimation of the quantities under study. Though several solutions have been proposed to import complex geometries into Geant4, so far, only commercial options are available. In this paper we present a new tool, GUIMesh, that embeds FreeCAD libraries, an open-source CAD editor, to tessellate volumes, and convert them to Geometry Description Markup Language (GDML), a Geant4 readable format, in a straightforward way. Several degrees of freedom are given to the user regarding the mesh and choice of material. Different geometries were tested for p material definition, geometry and navigation errors, successfully validating the method used.

**1. Introduction**

Particle transport simulation is fundamental for space and medical applications as well as for nuclear and particle physics. Geant4 is a C++ object-oriented framework developed to simulate particle interactions and track their path in materials [1][2][3]. Geometries in Geant4 can be described through several innate classes, which reproduce simple geometric shapes such as parallelepipeds, spheres and cylinders amongst others, and combining them via Boolean operations. For more irregular solids, tessellated surfaces can also be used to define geometries. This can be done using Geant4 native class G4Tesselation or by a Geometry Description Markup Language (GDML) [4][5] file, a geometry description format based on XML. The CADMesh [6][7] tool already allows to interface Stereolithography (STL is a CAD mesh format) [8] files with Geant4. However, there are currently no non-commercial tools capable of interfacing STEP [9] described geometries, the most widely used data exchange CAD format, to Geant4 by converting them to GDML.

In this work FreeCAD [10], an open-source CAD editor with Python [11] scripting capabilities, was used to tessellate volumes and produce GDML files to be input to Geant4. Since FreeCAD libraries can be easily imported to Python, a Graphical User Interface (GUI) was programed using Python and resorting to FreeCAD libraries capabilities to handle STEP files and mesh different geometries into a tessellated solid. This application, GUIMesh, allows users to import STEP files using FreeCAD properties, manage materials - STEP files do not include material information-, mesh volumes, and export them to GDML files. Materials are chosen from the Geant4 material database based on the NIST [12] library or created by mixing materials listed in that database. Assignment of materials to volumes can be performed on a one-to-one basis or loaded from a text file. A default material is assigned to loaded volumes. Meshing is performed via FreeCADs "tessellate" function which converts all surfaces into a mesh of triangles. For each volume a GDML file is written to provide some versatility. An extra GDML file is also written, the "mother" GDML, which is read by the GDML parser function in Geant4, providing the geometry tree so that multiple volumes can be integrated in the framework.

## 2. Methods

GUIMesh can be divided in three functionalities: CAD processing using FreeCAD to read step files and tessellate volumes, material definition and assignment to volumes and GDML file writing.

### 2.1. Interface

The Graphical User Interface (GUI) built with Tkinter, a Python extension, to facilitate the procedure of converting CAD geometries to GDML, is shown in Fig. 1. Three parts can be distinguished in it. On the left side, a menu with eight buttons, corresponding to different user options can be seen. The middle section shows the volume list of the imported STEP file as well as a list with the assigned materials. By default, silicon is assigned to all volumes when a file is first opened. On the right side there is a panel with the editing options. There, volume material and density can be changed by the user as well as the maximum meshing surface deviation. The material database is also displayed. New materials can be added to it as discussed in section 2.3.

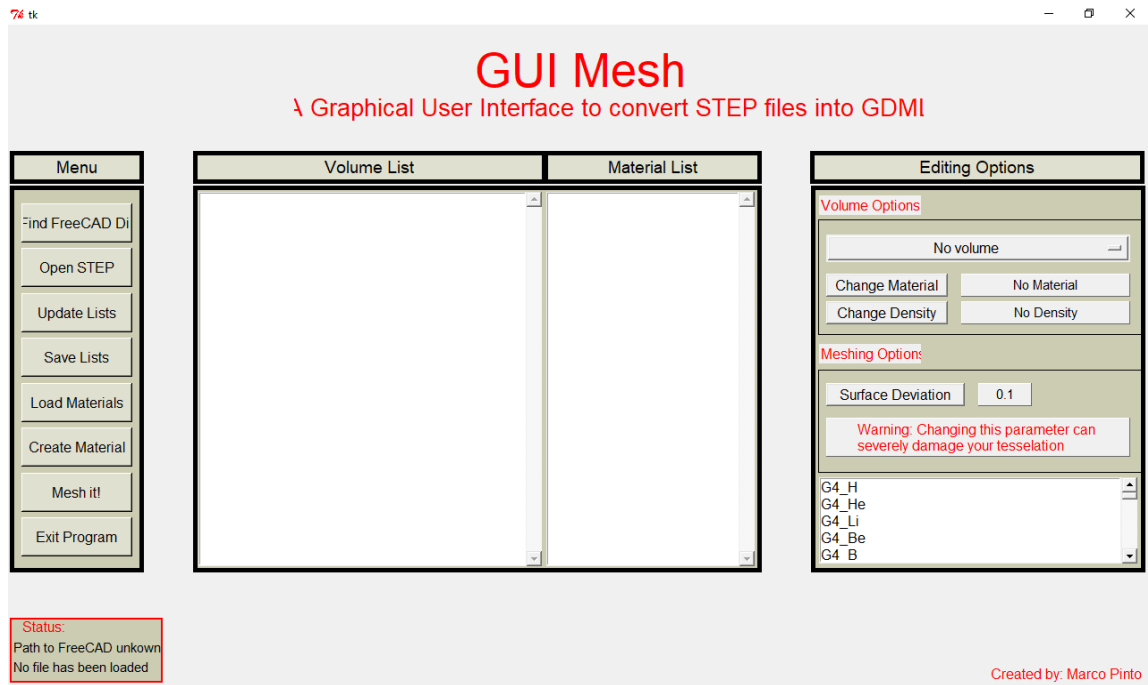

Fig. 1 GUIMesh GUI. On the left side a menu with different functions is presented to the user as buttons. The middle panel displays the list of the loaded volumes and their assigned materials. The right side is dedicated to defining volume and meshing options.

### 2.2. CAD Processing

GUIMesh imports FreeCAD libraries with Python 2.7[11]. Both 0.15 and 0.16 versions of FreeCAD are compatible with the GUIMesh though volume name varies between versions. FreeCAD is used to import STEP files (.STEP and .STP extensions) via its "Import" module.

Tessellation of surfaces is performed by the FreeCAD standard meshing algorithm, which has one degree of freedom, the Maximum Mesh Deviation (MMD) with set at 0.1mm. MMD can be changed in GUIMesh: lower MMD values provide increased volume precision, but require larger memory allocation at simulation level, due to the increase in the number of triangles. Precision for planar surfaces does not change with MMD.

### 2.3. Material Definition

All (270) materials from the NIST library are predefined in GUIMesh and can be accessed directly using their Name in the Geant4 Material Data Base (G4_H for Hydrogen, G4_He for Helium etc.). These materials cannot be changed but they can be mixed to create new materials of a given name, density and number of elements, each representing a fraction of the material composition. When loaded into GUIMesh all volumes are given a default material, Silicon (G4_Si), that can be altered directly in GUIMesh or latter in Geant4. In the former case, materials can be assigned to volumes on a one-by-one basis, or uploaded through a text file containing all materials, where each line contains the material name assigned to the volume with the same number.

*2.4. GDML*

GDML is an XML-based language interchangeable format, designed to describe geometries for physics simulations. This format accepts both elementary shapes and tessellated solids. Tessellation was the method chosen for GUIMesh, since it allows do describe complex irregular solid [4][5].

There are five fields accepted by the GDML format: define; materials; solids; structure and setup.

The field "define" is used to define different values that may be used in the rest of the file. Here is where the vertex positions of the volume in consideration are registered. An example, consisting on the positions of the vertices of a cube, is given below:

```
<define>
        <position name="Cube1_v0" unit="mm" x="0.0" y="21.0" z="0.0"/>
        <position name="Cube1_v1" unit="mm" x="0.0" y="11.0" z="0.0"/>
        <position name="Cube1_v2" unit="mm" x="0.0" y="11.0" z="10.0"/>
        <position name="Cube1_v3" unit="mm" x="0.0" y="21.0" z="10.0"/>
        <position name="Cube1_v4" unit="mm" x="10.0" y="11.0" z="10.0"/>
        <position name="Cube1_v5" unit="mm" x="10.0" y="11.0" z="0.0"/>
        <position name="Cube1_v6" unit="mm" x="10.0" y="21.0" z="0.0"/>
        <position name="Cube1_v7" unit="mm" x="10.0" y="21.0" z="10.0"/>
 </define>
```

The field "materials" is used to define the materials assigned to the objects. NIST materials do not need to be defined in detail, since Geant4 will correctly interpret the material name if it corresponds to the Geant4 Material Database reference. Any compound or mixture of materials however, may be defined in this field as a mix of NIST materials, each defined by the corresponding mass fraction in the compound or mixture. An example is given below:

```
<materials>
        <material formula=" " name="Water" >
                    <D value="1.00 " unit="g/cm3"/>
              <fraction n="0.667" ref="G4_H" />
              <fraction n="0.333" ref="G4_O" />
      </material>
</materials>
```

The field "solids" corresponds to the geometric definition of volumes. The world volume is defined as a simple box in the mother file. Tessellated solids from the STEP file are registered as a series of triangles enclosing a surface. Below, an example that defines the triangles based on the vertex positions of the cube defined in field "define", is shown:

```
<solids>
        <tessellated aunit="deg" lunit="mm" name="Cube_solid">
                <triangular vertex1="Cube_v0" vertex2="Cube_v1" vertex3="Cube_v2"/>
                <triangular vertex1="Cube_v3" vertex2="Cube_v0" vertex3="Cube_v2"/>
                <triangular vertex1="Cube_v4" vertex2="Cube_v5" vertex3="Cube_v6"/>
                <triangular vertex1="Cube_v4" vertex2="Cube_v6" vertex3="Cube_v7"/>
                <triangular vertex1="Cube_v1" vertex2="Cube_v5" vertex3="Cube_v4"/>
                <triangular vertex1="Cube_v2" vertex2="Cube_v1" vertex3="Cube_v4"/>
                <triangular vertex1="Cube_v7" vertex2="Cube_v6" vertex3="Cube_v0"/>
                <triangular vertex1="Cube_v7" vertex2="Cube_v0" vertex3="Cube_v3"/>
```

```xml
            <triangular vertex1="Cube_v1" vertex2="Cube_v0" vertex3="Cube_v6"/>
            <triangular vertex1="Cube_v5" vertex2="Cube_v1" vertex3="Cube_v6"/>
            <triangular vertex1="Cube_v7" vertex2="Cube_v3" vertex3="Cube_v2"/>
            <triangular vertex1="Cube_v7" vertex2="Cube_v2" vertex3="Cube_v4"/>
        </tessellated>
 </solids>
```

The field "structure" defines the geometry hierarchy. Since we opted to write one file for each volume, it only describes the solid and material of the specified volume:

```xml
<structure>
        <volume name="Cube">
            <materialref ref="G4_H"/>
            <solidref ref="Cube_solid"/>
        </volume>
 </structure>
```

The hierarchy is defined by inserting all volumes into the world volume, as in the e following "structure" field definition:

```xml
<structure>
        <volume name="World">
            <materialref ref="Vacuum_ref"/>
            <solidref ref="WorldBox"/>
                <physvol>
                <file name="Cube_1.gdml"/>
                <positionref ref="center"/>
                <rotationref ref="identity"/>
                </physvol>
                        .
                        .
                        .
                <physvol>
                <file name="Cube_N.gdml"/>
                <positionref ref="center"/>
                <rotationref ref="identity"/>
                </physvol>
        </volume>
</structure>
```

There is no limit to the number of physical volumes that can be assigned to the global geometry with this method. Finally, the field "setup" identifies which is the top volume. For each GDML file, the top volume corresponds to the volume defined in the field "solids". For the mother GDML file, the top volume is world volume.

The geometry thus defined can then be imported into Geant4 by the GDMLParser function [5], with the mother.gdml file as argument.

### 2.5. Geometry Validation

To confirm that the volumes are correctly imported into Geant4, navigation and material tests were performed. Three test geometries were defined: a cylinder with a 5mm radius and 10mm height; a torus with 10mm outer radius and 5mm radius section; and a 5mm sphere. These were tested with MMD=0.1mm, since due to their curved surfaces they are more prone to precision problems than plane geometries. Since we expect to import several volumes at once into Geant4, a system of two spheres, where a larger 5mm radius sphere, with a 2mm radius spherical hole in its center, encases a second sphere with the same size as the hole, meshed with different MMDs was also tested. Fig. 2 shows all tested solids in STEP format visualized with FreeCAD.

All solids imported into Geant4 were visually analysed for correctness. The precision of the system with two spheres was also analysed with the Geant4 test particle, the geantino, which is a "particle" that crosses the geometry and does not interact with any materials. 10E+4 Geantinos were omnidirectionally generated from point-like source located in the center of the sphere system. The positions where the geantinos changed volumes, from the inner sphere to the outer sphere, and from the outer sphere to the outside world, were recorded. Since the spheres are meshed, the radius of the sphere is not constant in all directions. The average and maximum values of the radii of both spheres for different MMD were thus computed.

Navigation errors may occur in Geant4 due to particles getting stuck near triangle vertices, because their distance to two consecutive volumes is within the step length precision A particle is considered stuck in the geometry when it does not change position for 10 consecutive steps. When this happens Geant4 slightly moves the particle by 10E-7mm, aborting the simulation of that specific particle if the particle remains in the same position after the fix. To account for navigation errors 10E+4 geantinos were generated from an 8mm radius spherical source surrounding the test volumes. The number of particles stuck between volumes was registered as a fraction of the total number of particles. All solids presented in Fig. 2 were submitted to this navigation error test.

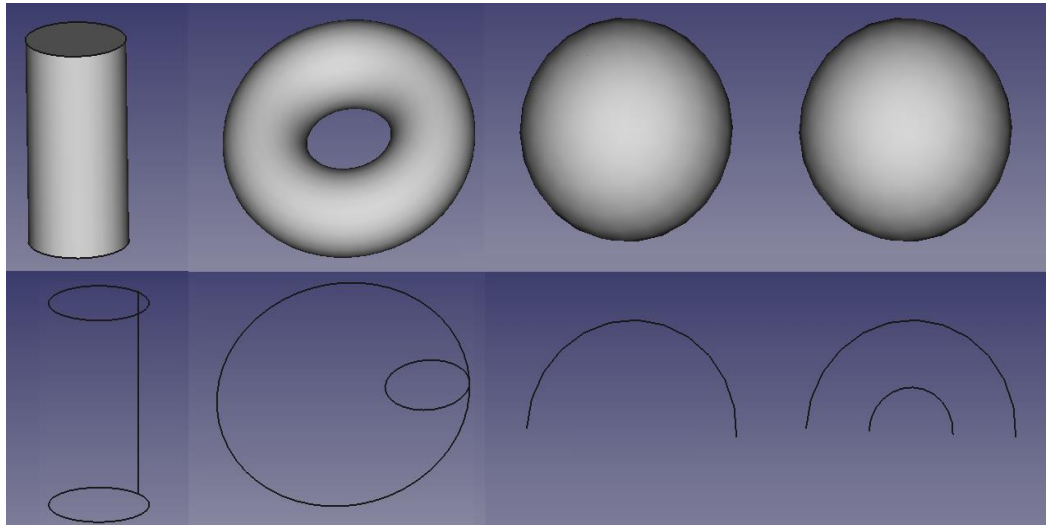

Fig. 2 STEP format of all solids used in the tests, a cylinder, a torus, a sphere and a second sphere with a spherical hole in the middle to encapsulate another a sphere that fills it. Below each solid, a wired image of its volume can be found for clarity.

An additional test was performed to compare the distributions of the energy deposited in meshed and Constructive Solid Geometry, (CSG) spheres. This was done for each type of solid (meshed and CSG), generating a parallel beam of electrons with energies ranging from 100 keV to 500 MeV from a circular plane, with the same radius as the sphere (5 mm), 10 mm away from it, and registering the deposited energy in the volume.

A summary of the solids, and of their characteristics, used for each test is shown in Table 1. The total surface area to MMD ratio is also given since it can be used as reference for the mesh precision for a given shape. Notice that it the precision is smaller in the 2mm radius sphere than in the 5mm radius sphere for the same MMD, since mesh deviation will be larger when compared to the total surface area.

Table 1 Summary of the solids used in the tests

| Solids | | | Tests | | | |
|---|---|---|---|---|---|---|
| Shape | Mesh Max Deviation (mm) | Area/MMD (#/mm$^2$) | Geometry | | Navigation | Energy Deposition |
| | | | Visual | Precision | | |
| Cylinder | 0.1 | 7.85E03 | X | - | X | - |
| Torus | 0.1 | 1.97E04 | X | - | X | - |
| Sphere | CSG | - | - | - | - | X |
| | 0.1 | 3.14 E03 | X | - | X | X |
| | 0.01 | 3.14 E04 | - | - | - | X |
| Sphere System Outer+Inner | 10 | 3.64E01+5.03E00 | X | X | X | - |
| | 1 | 3.64E02+5.03E01 | | | | |
| | 0.1 | 3.64E03+5.03E02 | | | | |
| | 0.01 | 3.64E04+5.03E03 | | | | |
| | 0.001 | 3.64E05+5.03E04 | | | | |

*2.6. Benchmarking*

To understand how tessellation affects simulation time and memory usage, simulations with the same configuration as the precision tests (1000000 geantinos), but without any readout, were performed for the Sphere System for the five MMD values mentioned in Table 1, as well as for a similar configuration, implemented with CSG solid class in Geant4. All simulations ran on a system with an Intel(R) Xeon(R) CPU E5540 @2.53GHz with 4 GB available memory.

## 3. Results

Several tests to the imported geometry and to particle tracking inside the volumes were performed to validate the method. CPU time and memory usage were also compared using the sphere system for different MMD values and for one defined by CSG with Geant4 user classes.

*3.1. Geometry*

All solids were successfully imported into Geant4. Fig. 3 shows the imported solids, where all single volume geometries were meshed with 0.1mm MMD. The two-sphere system was meshed with 10mm, 1.0mm, 0.1mm, 0.01mm and 0.001mm to understand the effect of this parameter. The shape of the spheres is highly dependent on the MMD. In fact, in Fig. 2 we can see that we can have a shape that no longer resembles the original sphere from Fig. 1 (10mm MMD) and good spherical approximations (0.1mm, 0.01mm and 0.001mm MMD). No overlaps were detected with [1][2][3].

The average and maximum deviation of the radius in the meshed spheres is given in Table 2. As expected both the average and maximum deviation become smaller for smaller MMD. It is also noticeable that the inner surface radius has a larger deviation than the outer one.

Table 2 Average and Maximum deviation from the mathematical surfaces of the sphere systems.

| MMD (mm) | Surface | Average Deviation (%) | Maximum Deviation (%) |
|---|---|---|---|
| 10 | Outer (5mm) | 41,4032 | 63,1302 |
| | Inner (2mm) | 41,403 | 63,1302 |
| 1 | Outer (5mm) | 5,0308 | 17,4846 |
| | Inner (2mm) | 11,5995 | 34,002 |
| 0,1 | Outer (5mm) | 0,8344 | 1,6554 |
| | Inner (2mm) | 1,7125 | 5,0685 |
| 0,01 | Outer (5mm) | 0,07 | 0,3052 |
| | Inner (2mm) | 0,166 | 0,7675 |
| 0,001 | Outer (5mm) | 0,0092 | 0,0326 |
| | Inner (2mm) | 0,02 | 0,0845 |

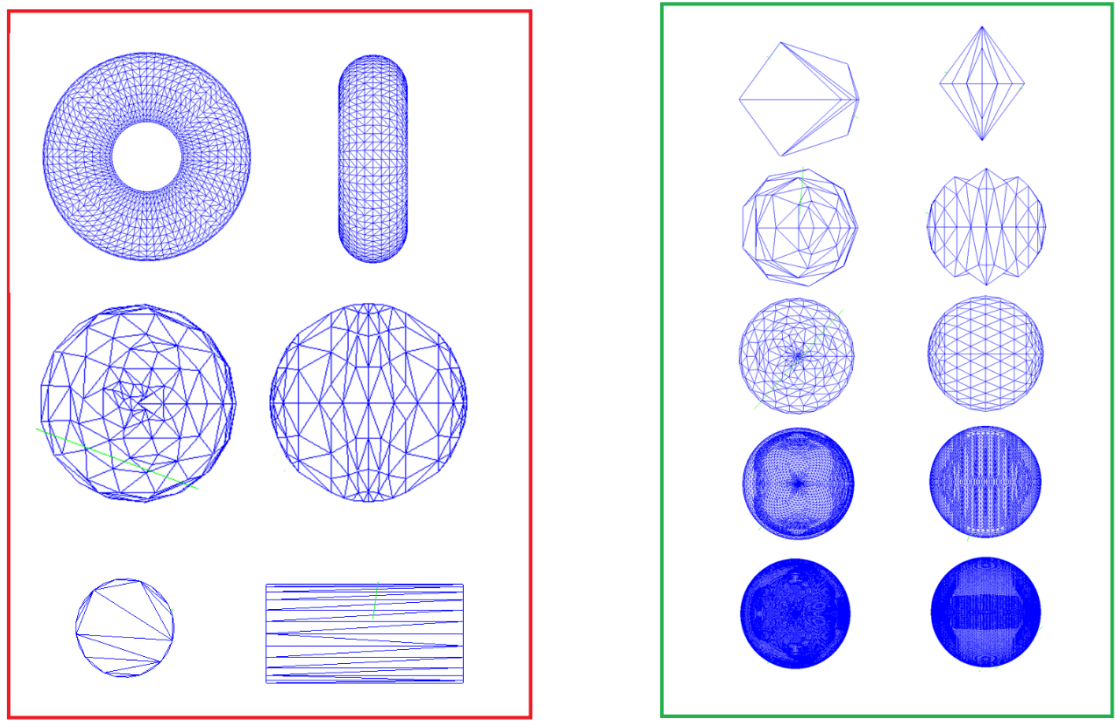

Fig. 3. Solids imported with Geant4. In the red box the single geometry solids mesh with 0.1mm MMD can be seen. In the green box the two-sphere system can be seen (only the outside sphere is visible) meshed with 10mm, 1mm, 0.1mm, 0.01mm and 0.001mm from top to bottom.

*3.2. Navigation*

The navigation test results are presented in Table 3. The aborted number of events with 0.1mm MMD is always smaller than 0.1%, varying with shape. For the sphere system. the number of aborted events is closely related to the MMD, decreasing proportionally with the MMD value. MMD.

Table 3 Navigation test results of all geometries. The aborted events percentage was calculated considering only particles that entered the test volume.

| Shape | MMD (mm) | Aborted Events (%) |
|---|---|---|
| Cylinder | 0.1 | 3,62E-04 |
| Sphere | 0.1 | 5,18E-04 |
| Torus | 0.1 | 6,89E-03 |
| Sphere System | 10 | 7,35E+00 |
| Sphere System | 1 | 4,62E-01 |
| Sphere System | 0.1 | 3,28E-02 |
| Sphere System | 0.01 | 2,86E-03 |
| Sphere System | 0.001 | 7,14E-04 |

*3.3. Deposited Energy*

The distributions of the energy deposited by electrons, with energies ranging from 100 keV to 500 MeV, in three 5mm radius spheres, two of them meshed with different precisions and a third defined using the Geant4 CSG sphere class, are shown in Fig. 4. The low precision meshed sphere (green line) shows different deposited energy when compared to the other two tested solids, although the total deposited energy differs from the two other cases by approximately 1%. The other two energy profiles, for the more precise meshing and for the CSG sphere implementation display an excellent agreement.

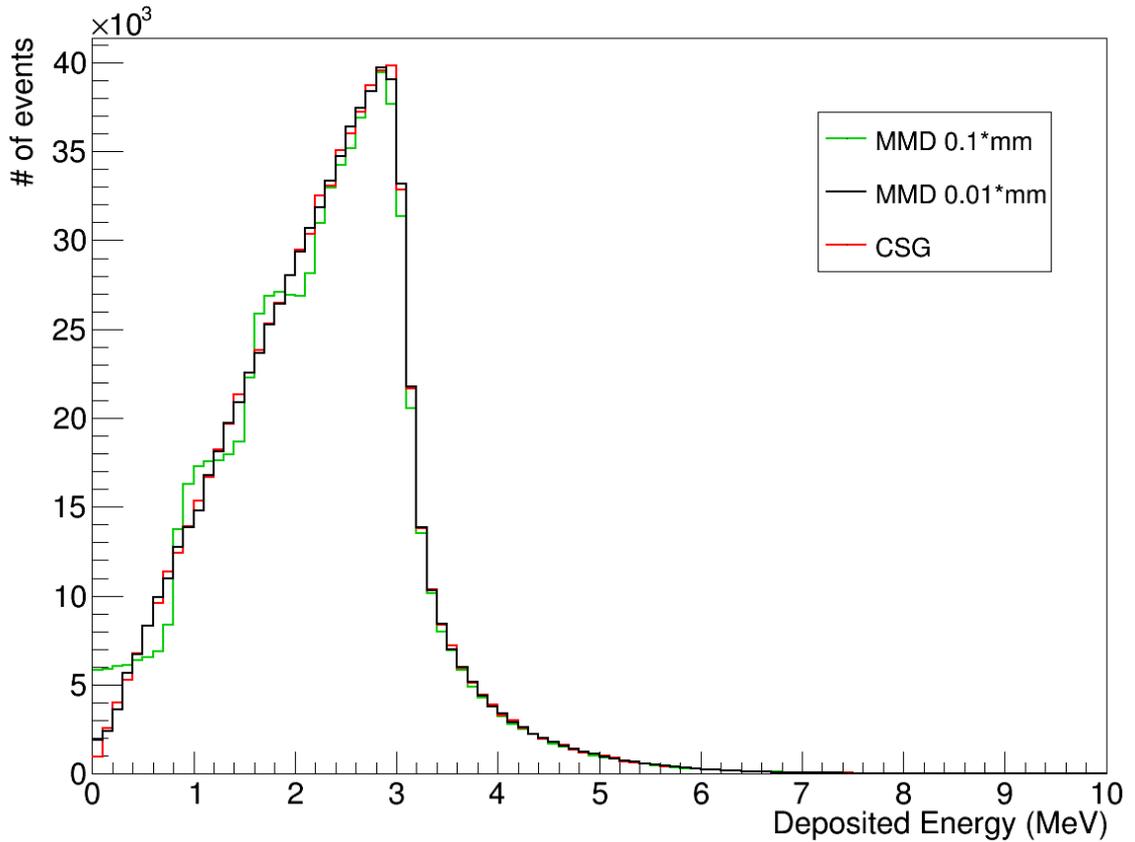

Fig. 4. Comparison between the deposited energy of electrons in three spheres, two meshed with different precisions and one implemented with a Geant4CSG class. Energy deposited in the 0.01*mm MMD sphere (black line) is similar to the deposited energy in the CSG sphere (red line). The energy spectrum in the 0.1*mm MMD sphere (green line) however is different due to its structural defects.

*3.4. Benchmarking*

Fig 5 shows computing time (left side) and memory used (right side) for 1000000 events runs, with different implementations of varying precision meshed geometries and for a CSG implementation, all with the same material, Silicon. Computing time increases exponentially with the precision of the mesh. Notice that CSG is always faster, even when compared to the sphere system meshed with the larger MMD value. Maximum Memory usage is dominated by system and Geant4 processes for all cases except for the highest precision meshing for which geometry accounts for ~33% of the memory.

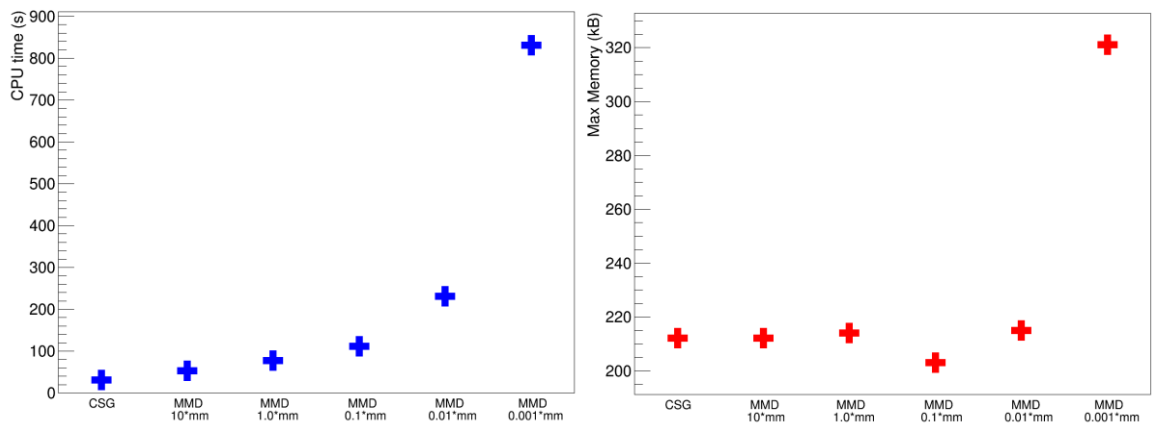

Fig. 5 Comparison of simulation CPU time (left side) and maximum memory usage (right side) between different mesh precision of the sphere system and CSG solid.

## 4. Discussion

Several geometries in STEP format were successfully imported into Geant4 via GUIMesh. No defects (holes) were found in the tessellated geometries. For curved surfaces shape form was shown to be highly dependent on the value of MMD. This is due to MMD limiting the maximum distance from an edge of the mesh to the original surface that it describes, e.g., larger MMD values allow larger deviations from this surface. This results in finer tuning of the surfaces for lower MMD values, as the number of triangles increases and their size decreases. Deviation, defined as the distance from the surfaces in Geant4 to the solids mathematical definition, was also computed and shown to be smaller or of the same order of magnitude as the MMD value. This means that for the same MMD spheres with different radius will have different normalized deviations hence when choosing the MMD one should also consider the size of the solids to be meshed. No overlaps were detected since all triangles of the meshed solids are inside the original shape.

Navigation errors resulting from "stuck" particles and causing aborted events in some cases were studied. These happen when Geant4 is not able to compute in which volume the particle is located. The percentage of events displaying this behavior was under <0.1% for 0.1mm MMD solids, with increasingly smaller number of errors for lower MMD values. The distributions of deposited energy were found to depend on the meshing precision. In fact, while the distribution of energy deposited on a 0.1*mm MMD sphere displayed discontinuities, for the 0.01*mm MMD sphere results matched perfectly those obtained for the CSG sphere.

Memory and time consumption were found to increase for solids meshed with lower MMD values, which means that for complex and/or large number of solids careful consideration should be made regarding solid precision.

## 5. Conclusion

GUIMesh is framework enabling to import CAD geometries in STEP format into Geant4. A STEP file describing a global geometry is read by GUIMesh, the geometry is meshed with a given Maximum Mesh Deviation and registered with the corresponding material in a GDML file structure, readable by Geant4. A set of test volumes were used to test the GUIMesh performance in terms of geometrical accuracy particle navigation errors and material implementation It was concluded that geometry precision is highly dependent on the chosen value for MMD. Lower values of MMD result in more accurate geometries and less errors in particle tracking, at the cost of memory and processing time. Therefore, the values of MMD should be optimized for a given geometry considering both the required accuracy and the simulation processing time. should be optimized by the user, depending on the geometry.

No overlaps were found in the tested geometries. It was shown that material definition in GUIMesh successfully implements materials into Geant4.

GUIMesh can useful tool for all users of Geant4 allowing them to import STEP geometries of arbitrary size and complexity into the simulation toolkit. It can be obtained by contacting the authors.

## References

[1] Agostinelli, S.; et al., Nucl. Instum. Methods A 506, 3 (2003) 205.
[2] Allison, J.; et al., Geant4 developments and applications, IEEE Transactions on Nuclear Science Vol 53, 1, pp: 270-278, 2 FEB 2006.
[3] J. Allison et al, Recent Developments in Geant4; NIMA; Vol 835, Nov. 2017, pp 186-225
[4] R. Chytracek, J. McCormick, W. Pokorski, G. Santin ;Geometry Description Markup Language for Physics Simulation and Analysis Applications; IEEE Trans. Nucl. Sci., Vol. 53, Issue: 5, Part 2, 2892-2896
[5] http://gdml.web.cern.ch/GDML/doc/GDMLmanual.pdf
[6] Poole, C.; et al. , A CAD interface for GEANT4. Australasian Physical and Engineering Sciences in Medicine, 35(3), pp. 329-334, 2012.
[7] Poole, C. and Cornelius, I. and Trapp, J. and Langton, C.M.; Fast Tessellated Solid Navigation in GEANT4; IEEE Transactions on Nuclear Science Vol 99, pp: 1-7; 2012
[8] 3D Systems, Inc. Stereolithography Interface Specification, July 1988
[9] STEP-file, ISO 10303-21 -- Industrial automation systems and integration -- Product data representation and exchange -- Part 21: Implementation methods: Clear text encoding of the exchange structure
[10] https://www.freecadweb.org/

[11] www.python.org
[12] https://physics.nist.gov/cgi-bin/Star/compos.pl